\documentclass[%
prb, twocolumn, superscriptaddress
]{revtex4-2}

\usepackage{graphicx}
\usepackage{dcolumn}
\usepackage{bm}
\usepackage[dvipsnames]{xcolor} 

\usepackage{hhline}

\begin{document}


\title{\textbf{On the mechanism of ferromagnetic resonance in ferromagnet-superconductor trilayers}}

\author{Dariia Popadiuk}
 \affiliation{
V. G. Baryakhtar Institute of Magnetism of the National Academy of Sciences of Ukraine, Kyiv 03142, Ukraine
}%
\affiliation{
Nanostructure Physics, Royal Institute of Technology, 10691 Stockholm, Sweden
}%

\author{Julia Kharlan}
\email{yuliia.kharlan@amu.edu.pl}
 \affiliation{
V. G. Baryakhtar Institute of Magnetism of the National Academy of Sciences of Ukraine, Kyiv 03142, Ukraine
}%
\affiliation{
Institute of Spintronics and Quantum Information, Faculty of Physics, Adam Mickiewicz University, Pozna\'n, Uniwersytetu Pozna\'nskiego 2, 61-614 Pozna\'n, Poland
}

\author{Anatolii Kravets}
\affiliation{
Nanostructure Physics, Royal Institute of Technology, 10691 Stockholm, Sweden
}%
\affiliation{
V. G. Baryakhtar Institute of Magnetism of the National Academy of Sciences of Ukraine, Kyiv 03142, Ukraine
}%
\author{Vladislav Korenivski}
\affiliation{
 Nanostructure Physics, Royal Institute of Technology, 10691 Stockholm, Sweden
}%
\author{Jarosław W. Kłos}
\affiliation{
Institute of Spintronics and Quantum Information, Faculty of Physics, Adam Mickiewicz University, Pozna\'n, Uniwersytetu Pozna\'nskiego 2, 61-614 Pozna\'n, Poland
}%
\author{Vladimir Golub}
\affiliation{
V. G. Baryakhtar Institute of Magnetism of the National Academy of Sciences of Ukraine, Kyiv 03142, Ukraine
}%



\begin{abstract}
Temperature dependent magnetic properties of superconductor–ferromagnet–superconductor (SC/FM/SC) trilayers are studied both experimentally and theoretically, with a focus on ferromagnetic resonance (FMR). The influence of the SC and FM layer thicknesses on the FMR field is examined. To differentiate the mechanisms involved, we additionally investigate structures containing nonmagnetic metallic (M) or insulating (I) spacers (SC/FM/M/SC or SC/FM/I/SC). All the studied  multilayers show large reductions in the FMR field below the critical temperature of the SC, except the system containing an insulating spacer (SC/FM/I/SC). This SC-induced FMR-shift (resonance field/frequency) is larger for thicker SC as well as FM layers, reaching a saturation value for very large thicknesses. To explain the measured results, an analytical model is developed, in which the FM-magnetization precession modulates the magnetic flux in the system, thereby inducing an alternating supercurrent in the SC, which in turn produces a dynamic back-action magnetic field on the FM that shifts its resonance frequency. The model considers closed current loops, where the FM layer conductively links the supercurrents flowing in the opposite directions in the two outer SC layers. Our results provide a practical route for increasing the operating frequency of magnonic devices.

\end{abstract}

\maketitle


\section{\label{sec:level1}Introduction\protect}

Hybrid ferromagnet–superconductor (FM–SC) systems attract considerable attention because they enable control of the ferromagnet's magnetization dynamics \cite{yu2025} and spin currents \cite{Cai_2022} through electromagnetic or current-mediated coupling with the superconductor. The simplest realizations of such systems are multilayer structures, where FM layers are placed either in direct contact with SC layers or separated from them by an insulating spacer. The dynamic properties of such multilayers have been extensively studied, both experimentally \cite{Bell_2008, Jeon_2019, Golovchanskiy_2020, Golovchanskiy_2021a, Golovchanskiy_2023, Li2025EfficientFMR} and theoretically \cite{Volkov_2009, Mironov_2021, Silaev_2022, Ojajarvi_2022, Kuznetsov_2022, Silaev_2023, Zhou_2023, Zhou_2024, Derendorf_2024, Gordeeva_2025}. In such systems, the ferromagnetic resonance (FMR) frequency can be tuned by changing temperature, materials parameters or geometry the system, such as the FM and/or SC layer thicknesses. This opens the way toward a new class of magnonic devices where coupling with the superconducting component provides an additional tool for control of the magnetization dynamics.

The coupling mechanisms between FM and SC can, to some extent, resemble those taking place in FM/non-magnetic metal (NM) hybrids. A conductor placed in the vicinity of FM modifies the stray-field distribution, thereby changing the frequency of the magnetization precession. In FM/NM systems, this effect is absent for the FMR mode, but becomes relevant for propagating spin waves (SWs) having finite wave vectors \cite{Seshadri_1970}. Similar effects occur in FM/SC bilayers without galvanic contact, where electromagnetic coupling with the SC increases the frequency of the propagating SWs \cite{Borst_2023, Zhou_2024}. In trilayers where a dielectric FM layer is sandwiched between two SC layers or a dielectric spacer is placed between metallic FM and SC layers, strong hybridization can occur between the FM's uniform Kittel mode and the electromagnetic Swihart mode confined between the SC layers \cite{Swihart_1961, Golovchanskiy_2021a, Silaev_2023}. Such hybridization appears in the long-wavelength limit (millimeter scale for GHz frequencies). For $k=0$, the Kittel-derived mode softens toward zero frequency, while the Swihart-derived mode is observed at non-zero frequency\cite{Silaev_2023}.

Charge transport in a nanostructure provides another channel of coupling. Two FM electrodes can be coupled via an NM nanocontact through non-local exchange \cite{Tserkovnyak_2005}, although this mechanism becomes inefficient for extended NM spacers \cite{Tserkovnyak_2005}. Magnetic coupling via an SC nanocontact is also possible \cite{Deutscher_2000}, when the electrode separation is comparable to the SC coherence length. Conversely, two SC nanoelements can be coupled through a half-metallic FM \cite{Keizer_2006}. Moreover, coupling of SC layers in Josephson junctions via tunneling currents across FM barriers -- either insulating or metallic, with a nonconducting spacer -- has been considered~\cite{Volkov_2009, Derendorf_2024}. Since the tunneling supercurrent decays exponentially with barrier thickness, its influence on the FM magnetization dynamics becomes relevant only for ultrathin FM layers. 

The aim of this work is to use a comprehensive set of SC/FM multilayers to verify theoretical predictions and corroborate experimental observations of how the proximity of SC affects FMR.
To identify the main mechanism, we investigate the dependence of FMR frequency on both SC and FM thicknesses at different temperatures. The studied trilayers, composed of Nb/Py/Nb, are fabricated in direct contact. Using a simple theoretical approach, we demonstrate that dynamical coupling in SC/FM/SC systems can be mediated by closed-loop currents. Such current loop is formed by the top and bottom SC layers, with the current in SC's flowing in the opposite directions (in-plane) and perpendicular to the plane of the FM. The mechanism proposed captures the interplay of the electro-magneto-dynamic effects in the system. We show that breaking of the screening-current loop, e.g. by spacing FM from SC with a thin insulator, suppresses the effect entirely. The results of our theoretical model are consistent with all of the experimentally observed trends and visualize well the physical mechanism at play.

The paper is organized as follows. In Sec. II (Experiment), we describe the samples studied and the experimental methods employed. Section III (Results and Discussion) presents the experimental results together with the outcomes of the theoretical modeling. Sec. IV (Theory) derives the main result for the FMR frequency of the considered systems. Finally, Sec. V (Summary) concludes the paper.

\section{\label{sec:level1}Experiment}

\subsection{\label{sec:level2}Materials}

\begin{figure}[b]
\includegraphics[width=0.9\columnwidth] {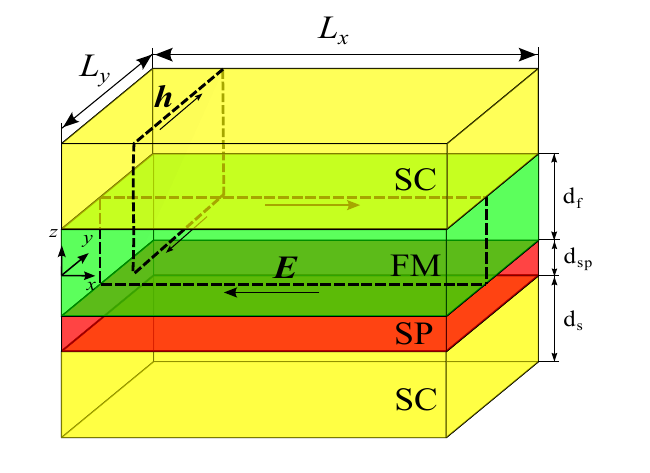}
\caption{\label{fig:SFS} Schematic of the multilayer structure of length $L$, consisting of a ferromagnetic Py layer of thickness $d_\mathrm{f}$, an optional spacer layer of thickness $d_{\mathrm{sp}}$, enclosed by two superconducting Nb layers of thickness $d_\mathrm{s}$. }
\end{figure}

A series of Nb/Py/Nb multilayers with varying layer thicknesses (see Table~\ref{tab:samples}) was deposited onto commercial Si/SiO$_2$ substrates (1 µm thermal oxide) at room temperature using UHV dc magnetron sputtering (multi-target Orion system, AJA Inc.). The films were grown in 3 mTorr Ar atmosphere. Prior to deposition, the substrates were pre-cleaned by Ar ion milling in 5 mTorr Ar at 45 W RF bias for 5 min. The thickness of the Nb and Py layers was controlled by deposition time, calibrated independently by profilometry.

In samples S1–S8, Nb and Py are in direct contact. In sample S9, a conducting Ag spacer was introduced between Py and Nb on one side, while in sample S10 insulating $\rm Al_2O_3$ spacer was placed on one side of Py. The $\rm Al_2O_3$ layer was deposited by reactive sputtering from a metallic Al target (dc magnetron) without breaking vacuum, ensuring high-quality interfaces. The insulating properties were confirmed by optical and transport measurements. For thicknesses above 1~nm, the oxide spacers are pinhole-free (verified in via magnetoresistance). For thicknesses of 2~nm and greater, the conductivity of the oxide barriers is negligible, and at 5 nm interlayer transport effects or other proximity effects can be disregarded. The conducting Ag spacer of thickness 2~nm was deposited from a metallic Ag target (dc magnetron).

\begin{table}[h]
\centering
\begin{tabular}{|l|l|l|l|l|}
\hline 
Sample & $d_{\rm f}$~(nm)  & $d_{\rm s}$~(nm) & $d_{\rm sp}$  (nm) & $H_{\rm a}$ (Oe) \\  [0.25ex]
\hhline{|=|=|=|=|=|}
S1 & 10 & 50 & none & -1332 \\ \hline
S2 & 20 & 50 & none & -767 \\ \hline
S3 & 50 & 50 & none & -661 \\ \hline
S4 & 100 & 50 & none & -519 \\ \hline
S5 & 200 & 50 & none & -150 \\ \hline
S6 & 500 & 50 & none & -36 \\ \hline
S7 & 50 & 100 & none & -75 \\ \hline
S8 & 50 & 200 & none & -626 \\ \hline
S9 & 50 & 50 & 2 (Ag) & -1146 \\ \hline
S10 & 50 & 50 & 5 ($\rm Al_2 O_3$) & -904\\ \hline
\end{tabular}
\caption{Geometrical parameters of the samples: thicknesses of the ferromagnetic layer ($d_\mathrm{f}$), superconducting layer ($d_\mathrm{s}$), and spacer ($d_{\mathrm{sp}}$). Material parameter: surface magnetic anisotropy of Py ($H_\mathrm{a}$).
}\label{tab:samples}
\end{table}

\subsection{\label{sec:level2}Measurements}

The technique used to investigate the dynamics in the system was ferromagnetic resonance (FMR), which provided direct information on the magnetization dynamics. To complement the FMR data with static characteristics of the FM material, we additionally performed magnetometry measurements.

FMR was measured at fixed frequency of 12 GHz using a coplanar waveguide–based spectrometer, mounted in a PPMS dewar enabling measurements down to 3~K in temperature and upto 9~T in dc magnetic field. The field was applied in the film plane (along the $x$-axis in Fig.\ref{fig:SFS}) and was swept through the resonance condition to record the spectra. The microwave excitation field, with an amplitude of about 1 Oe, was applied perpendicular to the dc field in the film plane (along the $y$-axis in Fig.\ref{fig:SFS}).

To assess whether the observed effect on FMR is purely dynamic, static magnetic characterization was carried out using a Vibrating Sample Magnetometer (VSM, DynaCool PPMS by Quantum Design).

The ntrinsic parameters of permalloy (Py) namely the saturation magnetization, gyromagnetic ratio, and surface anisotropy were determined from out-of-plane angular dependences of the resonance field at room temperature using a Bruker ELEXSYS E500 electron spin resonance cavity spectrometer operating in the X-band frequency range (9.86~GHz) and equipped with an automatic goniometer. These parameters show weak temperature dependence and were used in calculations of the low temperature properties.

Together, the above complementary magnetic measurement techniques provide a comprehensive characterization of the system and form the basis for the results discussed below.

\section{\label{sec:level1}Results and discussion}

\begin{figure}[h]
\includegraphics[width=1.0\columnwidth]{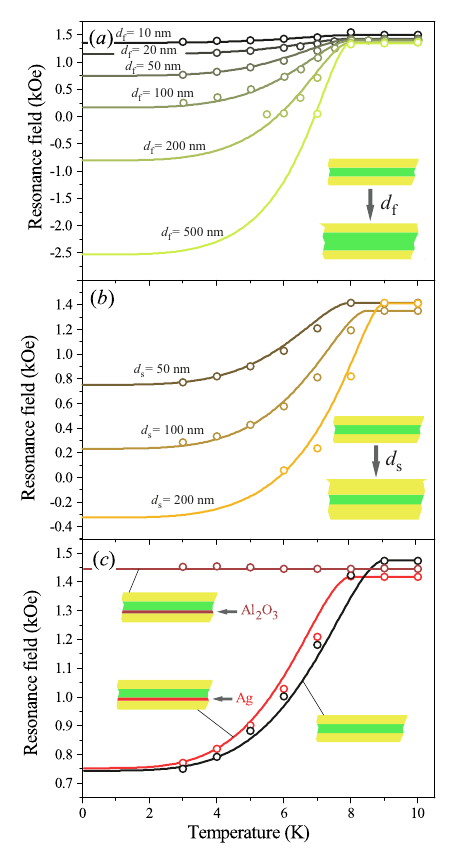}
\caption{\label{fig:FMR} Temperature dependence of the resonance field for the samples with: (a) $d_{\rm s} = 50$ nm and different values of $d_{\rm f}$; (b) $d_{\rm f} = 50$ nm and different values of $d_{\rm s}$; and (c) $d_{\rm f} = 50$ nm, $d_{\rm s} = 50$ nm, with different spacers (conductive: Ag and insulating: $\rm Al_2O_3$) between SC and FM. The resonance frequency is 12 GHz. Points and solid lines correspond to experimental data and theoretical calculations, respectively.}
\end{figure}

The data on FMR in SC/FM/SC multilayers are summarixed in Fig.~\ref{fig:FMR}. The points are the measured values of the FMR field at a fixed frequency of 12 GHz. The lines drown trough the data are theoretical fits sing the model developed in Sec.~\ref{sec:theory}.

Figure~\ref{fig:FMR}(a) shows the temperature dependence of the FMR field for samples S1–S6 for a fixed SC layer thickness of $d_{\rm s} = 50$ nm and the FM layer thicknesses of $d_{\rm f} = 10, 20, 50, 100, 200,$ and $500$ nm. In these samples, the Nb and Py layers are in direct contact, i.e., the M/I spacer thickness is zero: $d_{\rm sp} = 0$. Strong changes in FMR are observed at temperatures below the superconducting transition temperature ($T_\mathrm{C}$ of Nb): the resonance field drops sharply and tends to saturate as $T \to 0$. Moreover, the reduction in the resonance field becomes more pronounced with increasing FM layer thickness $d_{\rm f}$. This effect is absent in single FM layers and SC/FM bilayers, where the FMR field is practically independent of the layer thickness. It is also worth noting that for $T > T_{\rm C}$, the FMR field is practically unaffected by changes in $d_{\rm f}$.

The influence of the SC layer thickness is illustrated in Fig.~\ref{fig:FMR}(b), showing the temperature evolution of the resonance field for samples S3, S7, and S8 with $d_{\rm f} = 50$ nm, $d_{\rm s} = 50, 100,$ and $200$ nm, and $d_{\rm sp} = 0$. Similar to the case of thicker FM layer, the resonance field decreases more rapidly for thicker SC. At the same time, the superconducting transition temperature slightly increases with $d_{\rm s}$, reaching $T_{\rm C} = $8, 8.5, and 9~K for $d_{\rm s} =$ 50, 100, and 200~nm, respectively, which correlates with the results obtained previously~\cite{Li, Jeon_2019, Golovchanskiy_2020, Golovchanskiy_2023, Li2025EfficientFMR} for similar structures.

Finally, to clarify the role of inter-layer conductivity, we have investigated two extra samples having spacers between the SC and FM layers. Samples S3, S9, and S10 have the same thicknessses of FM and SC layers: $d_{\rm f} = 50$ nm and $d_{\rm s} = 50$~nm. In S3, the SC and FM layers are in direct contact; in S9 they are separated on one side by a metallic (M) spacer: Ag, of thickness $d_{\rm sp} = 2$~nm; and in S10 by an insulating (I) spacer: Al\textsubscript{2}O\textsubscript{3}, of thickness $d_{\rm sp} = 5$~nm. As shown in Fig.~\ref{fig:FMR}(c), the conducting spacer practically does no affect the FMR, whereas the insulating spacer completely suppresses the SC-indiced changes of the resonance field. This is a clear demonstration that inter-layer conduction (normal electrical current across the middle FM layer) is a necessary condition for the observed effect. In other words, the effect cannot be explained either by displacement or tunneling currents across the FM layer, since both mechanisms should still be operational in the presence of the thin I spacer.

\begin{figure}
\includegraphics[width=1.0\columnwidth]{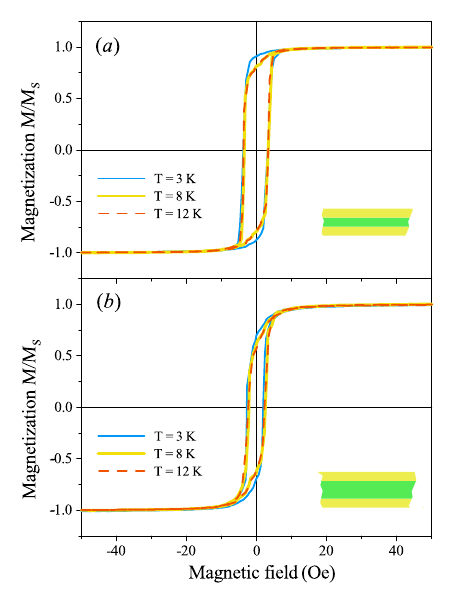}
\caption{\label{fig:hyst} Hysteresis loop measured for the sample S4 with $d_{\rm f}=100$ nm, $d_{\rm s}=50$ nm (a) and for the sample S5 with $d_{\rm f}=200$ nm, $d_{\rm s}=50$ nm (b).}
\end{figure}

To examine the influence of the effect of stray fields, which is considered to be a possible mechanism of the resonance frequency shift in such FM/SC systems~\cite{Mironov_2021}, quasi-static measurements of hysteresis loops at different temperatures for two samples with different FM layer thicknesses have been carried out. It was predicted in \cite{Mironov_2021} that the SC films can screen the static stray field of the FM layer, resulting in a giant demagnetization effect, with a corresponding shift of the resonance frequency. In our experiment, the magnetic field was applied in the film plane, and the magnetization was measured along the field direction. Normalized hysteresis loops for samples S4 and S5 were measured at $T = 3$, 8, and 12 K, and are shown in Fig.~\ref{fig:hyst}.

If the Nb films in the superconducting state generated static stray fields in the FM region due to the Meissner effect, one would expect a broadening of the hysteresis loops below the critical temperature. However, as it is clear from Fig.~\ref{fig:hyst}, the loops measured at 3 K (strongly superconducting state), 8 K (transition region), and 12 K (normal state) are nearly identical. Minor differences between the loops at 3 K and higher temperatures can be attributed to edge effects, but these changes of a few Oe are negligibly small on the scale of the observed FMR field changes of the order of kOe. This indicates that the decrease of the resonance field at low temperatures cannot associated with static stray fields and has a dynamic origin.

Introducing an insulating spacer between the FM and SC layers (blue points and line in Fig.~\ref{fig:FMR}(c)) leads to vanishing of the frequency shift effect. This correlates with earlier experimental observations reported in \cite{Li2025EfficientFMR}. It should be pointed out this result does not contradict the observed in \cite{Golovchanskiy_2021a} effect of frequency increase for the SC/FM/I/SC structure. In \cite{Golovchanskiy_2021a} the SC/FM/I structure was placed directly on top of the central transmission line of the SC waveguide used, which allowed Swihart mode exitation in the dielectric layer by the $ac$ current flowing through this line. This mode hybridized with the FMR mode, which can lead to the observed frequency upshift. So the mechanism proposed in \cite{Silaev_2023} is likely valid not for arbitrary SC/FM/I/SC or SC/FM/SC systems, but only for those where the Swihart mode can be externally excited. We cannot apply to our system the theoretical model developed by Zhou \cite{Zhou_2023}, which is based on the assumption that the SC reflects the radio-frequency electric field induced by the precessing magnetization in the FM insulator. According to this model, the SC currents oscillate in the top and bottom layers with a $\pi$-phase shift between them, and there is no current flow between the SC layers. The model assumes uniformity in the direction tangential to the interfaces and unconstrained in-plane dimensions. These seemingly obvious assumptions may not hold on the experiment, where the sample has finite dimensions. A current oscillating at GHz frequencies in a finite layer generates a non-uniform charge concentration compared to the current dynamics. The inhomogeneous charge distribution produces, in turn, an additional electric field that results in weakening of the current oscillations. This leads to the disappearance of the dynamic influence of the SC on precessing magnetization in the FM. It should be also mentioned that the displacement current, which can flow across the insulator, is extremely small and cannot practically affect the dynamics of the SC currents —- it is several orders of magnitude smaller than the dissipative (ohmic) current in the case of a conductive FM layer being in direct contact with the SC.

The analysis by Silaev \cite{Silaev_2022} of the dynamic magnetic field for infinitely extended SC/FM/SC trilayer with a conductive FM component cannot be used to explain the disappearance of the FMR field/frequency shift when the electrical conduction between the SC layers is blocked, since it assumes that currents in the SCs oscillate in the in-plane direction rather than circulating along a closed loop through the FM layer. Moreover, the model is not applicable to realistic finite-size samples: here, charge displacement in the SCs generates an internal electric field that compensates the external field and consequently drives the current to zero on a timescale of order $10^{-14}$ s, i.e., about four orders of magnitude shorter than the characteristic period of magnetization precession.

We believe it is sufficient to consider a simpler model, focusing on the analysis of currents in the SC/FM/SC system that can circulate in closed loops formed by the two SC layers connected by the conductive FM layer. Such a non-local approach is more intuitive and easier to interpret some of the key experimental findings. For example, it explains the absence of the observed FMR field/frequency shift when the SC layers are electrically isolated from each other, or the change in FMR field/frequency with the changes in FM and/or SC layer thickness.

The main idea behind the proposed model can be outlined as follows. In SC/FM/SC systems, the external magnetic field creates a torque that induces magnetization precession, leading to a periodic change in the magnetic flux in the system. According to Faraday’s law, this flux results in the generation of an alternating electric current in the SC, which in turn produces an alternating magnetic field in the FM film, thereby shifting the FMR frequency. Importantly, such a current can arise only in the case of a conducting FM without a dielectric spacer, since only then is the electric loop encompassing the FM closed.

The derivation of the FMR frequency is presented in detail in Sec.~\ref{sec:theory}. Here, we focus only on the final expression for the FMR frequency:
\begin{equation}
\label{ec:omega}
f=\frac{\gamma}{2\pi}\sqrt{(H_0-H^{\rm sc})(H_0+4\pi M_{\rm s}+H_{\rm a})},
\end{equation}
where $H_0$ is the in-plane external magnetic field, $M_{\rm s}$ is the saturation magnetization, $\gamma$ is the gyromagnetic ratio, and $H_{\rm a}$ is the perpendicular Néel-type anisotropy field. The correction $H^{\rm sc}$ resulting from the presence of the SC layers is given by
\begin{equation}
\label{ec:Hy_sc}
H^{\rm sc}=-4\pi M_{\rm s}\frac{d_{\rm f}\left[1 -\exp(-d_{\rm s}/\lambda) \right]}{2\lambda+d_{\rm f}\left[1 -\exp(-d_{\rm s}/\lambda) \right]},
\end{equation}
with $\lambda$ denoting the London penetration depth.

Material parameters such as the saturation magnetization, gyromagnetic ratio, and surface anisotropy for all samples S1–S10 were determined using the angle-dependent out-of-plane FMR measurements performed at room temperature, where the dynamic properties are similar to those of an isolated FM layer. Fitting the angular dependence of the resonance field to the Kittel formula yields $M_{\rm s} = 820$G and $\gamma = 3.03\times10^6$. The surface anisotropy $H_{\rm a}$ varied between the samples due to the differences in thickness and interfaces, as summarized in Table\ref{tab:samples}. The London penetration depth $\lambda$ decreases with increasing SC film thickness. The zero-temperature values $\lambda_0 = \lambda(T = 0)$ were taken as 113~nm, 108~nm, and 92~nm for samples S1–S8 with $d_{\rm s} = 50$~nm, 100~nm, and 200~nm, respectively. These values were used to fit the experimental data in Fig.\ref{fig:FMR} and are consistent with previously reported results\cite{Gubin}. For sample S9 with $d_{\rm s} = 50$~nm, $\lambda_0 = 106$nm, reflecting the effect of interfacing the SC with the Ag layer. The temperature dependence of $\lambda$ was calculated using the empirical law $\lambda=\lambda_0/(1-(T/T_{\rm c})^4)$\cite{Tinkham1996}.

Using Eqs.(\ref{ec:omega}) and (\ref{ec:Hy_sc}) with the parameters given above, we calculated the temperature dependence of the resonance fields. The results, shown in Fig.\ref{fig:FMR}, show a very good agreement between the theory (solid lines) and experiment (circles).

\begin{figure}
\includegraphics[width=1.0\columnwidth]{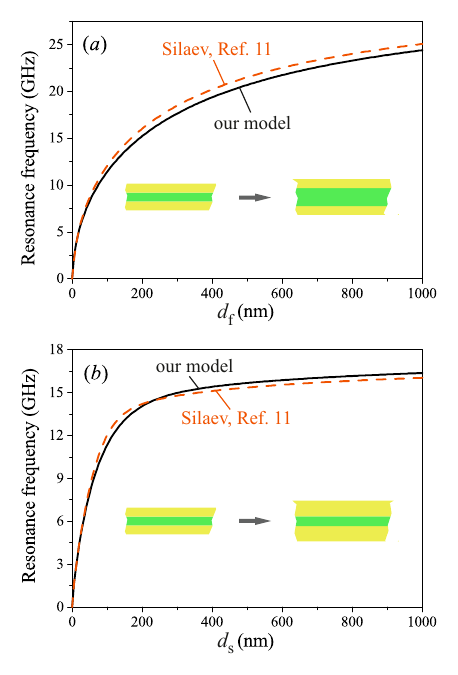}
\caption{\label{fig:comp} (a) Resonance frequency as a function of FM layer thickness $d_{\rm f}$ for $d_{\rm s}=50$~nm. (b) Resonance frequency as a function of SC layer thickness $d_{\rm s}$ for $d_{\rm f}=50$~nm. Both dependencies are plotted at $H_0=0$ and $T=0$. Black and red lines correspond to theoretical calculations based on our model and that of Silaev, respectively.}
\end{figure}

Having verified the theory versus experiment for the FMR shift, we analyzed the effect of the system’s geometric parameters on the resonance frequency. The black lines in Fig.\ref{fig:comp} show the theoretical dependencies calculated from Eqs.(\ref{ec:omega}) and (\ref{ec:Hy_sc}). Figure~\ref{fig:comp}(a) illustrates the variation with FM-layer thickness $d_{\rm f}$ at fixed $d_{\rm s}=50\ \mathrm{nm}$, while Fig.\ref{fig:comp}(b) shows the dependence on the SC-layer thickness $d_{\rm s}$ at fixed $d_{\rm f}=50\ \mathrm{nm}$. The thickness dependence of the London penetration depth $\lambda$ was taken from Ref.\cite{Gubin}. All calculations were performed at $T=0$. Since the anisotropy field scales as $H_{\rm a}= C/d_{\rm f}$, the values for samples S1–S6 (Table~\ref{tab:samples}) were used to fit this dependence for $C =- 21\times 10^4 {\rm \; Oe\;cm}$, and interpolate $H_{\rm a}$ for arbitrary $d_{\rm f}$.

The resonance frequency increases with increasing the FM film thickness because a larger cross-section produces more ac magnetic flux emanating from the precessing magnetization. However, the FMR frequency eventually saturates, as $H_y^{\rm sc} \to -4\pi M_{\rm s}$ for $d_{\rm f} \to \infty$. This saturation is due to Lenz’s law: superconducting currents induced in the SC oppose the flux change by reducing the magnetization precession, which in turn weakens the currents, leading to a self-consistent saturation of the frequency. Increasing the SC thickness also enhances the resonance frequency, since for $d_{\rm s}\gg 2\lambda$ the dynamic field is more effectively screened. In this limit, $H_y \to -4\pi M_{\rm s} d_{\rm f}/(2\lambda+d_{\rm f})$ as $d_{\rm s} \to \infty$, and the frequency again saturates. This behavior reflects the Meissner effect: magnetic and electric fields decay inside the SC, so for $d_{\rm s} \gg \lambda$ the region of the field penetration becomes independent of $d_{\rm s}$.

The red lines in Fig.~\ref{fig:comp} correspond to calculations based on the model of Silaev\cite{Silaev_2022}, which, interestingly, also show very good agreement with the experiment. Nevertheless, our model (Sec.~\ref{sec:theory}) explicitly incorporates features such as the current loops, which vanish in the limit of a uniform, infinitely extended layer but are essential for explaining the suppression of the effect when a dielectric spacer is introduced between the SC and FM layers. The model developed herein is relatively minimalist, yet it provides a robust explanatory framework for the multilayered material studied.

\section{\label{sec:theory}Theory}

In order to understand the experimentally observed strong increase (decrease) of the FMR frequency (field) in SC/FM/SC and SC/FM/M/SC structures (see Fig.~\ref{fig:SFS}) at low temperatures, and to contrast it with the behavior found in structures with insulation, we have developed an analytical model based on the concept of screening-current loops. The excited magnetization precesses around the direction of the static applied field $\mathbf{H}_0 = H_x \mathbf{e}_x$.

In the linear approximation, the $x$-component of the magnetization is constant and equal to the saturation magnetization $M_s$, while the small time-dependent components oscillate harmonically, $\mathbf{m}=(m_{y},m_{z})\,e^{i\omega t}$, and lie in the $yz$-plane. 

In this geometry, the FMR mode produces only the $z$-component of the dynamic demagnetization field, $h_z^{\rm d}=-4\pi m_z$, inside the FM, while the tangential component $h_y^{\rm d}$ vanishes. The total dynamic field $\mathbf{h}$, however, also contains a contribution $\mathbf{h}^{\rm sc}$ from the superconducting layers: $\mathbf{h}=\mathbf{h}^{\rm d}+\mathbf{h}^{\rm sc}$, where $\mathbf{h}^{\rm sc}$ has only a $y$-component, $h_y^{\rm sc}$. From the general relation $\mathbf{b}=\mathbf{h}+4\pi\mathbf{m}$, it follows that $b_z=0$ and $b_y=(h_y^{\rm sc}+4\pi m_{y})e^{i\omega t}$. Thus, the time-periodic flux associated with $b_y$, passing through any contour in the $xz$-plane, induces a circulating electric field. 

Let us consider a rectangular contour circulating around the FM layer in the $xz$-plane, with sides $L_x$ and $d_{\rm f}$ along the $x-$ and $z$-directions, respectively. This contour follows the boundaries of the FM film (see Fig.~\ref{fig:SFS}). According to Faraday's law, the change in magnetic flux leads to the appearance of an alternating electric field of non-zero curl:
\begin{equation}
\label{ec:Faradey_general}
\oint\mathbf{E}\,d\mathbf{l}=-\frac{1}{c}\int\frac{d\mathbf{B}}{dt}d\mathbf{S}. 
\end{equation} 
For the considered contour, Eq.~(\ref{ec:Faradey_general}) takes the form:
\begin{equation}
\label{ec:Faradey}
2E_0(L_x+d_{\rm f})e^{i\omega t}=-i\frac{\omega \,d_{\rm f}}{c}L_x (h_y^{\rm sc}+4\pi m_{y})e^{i\omega t}.
\end{equation} 
Taking into account that $L_x\gg d_{\rm f}$, the amplitude of the alternating electric field created on the FM boundary along the $x$-axis is
\begin{equation}
\label{ec:E0}
E_0=-i\frac{\omega d_{\rm f}}{2c}(h_y^{\rm sc}+4\pi m_{y})e^{i\omega t}.
\end{equation}
Due to the symmetry of the system, the electric field is oriented along the $x$-direction and has opposite orientation above and below the FM layer, as it circulates along the contour. 

In the case of an isolated FM film, $E_0$ would not depend on the distance from the FM, since the film is infinite. However, the presence of SC layers on both sides of the FM film leads to attenuation of the electric field along the $z$-direction in the SC region due to the Meissner screening effect, described by London theory. The penetration of the electric field into the SC layers can be found using the second London equation:
\begin{equation}
\label{ec:London2}
\frac{d^2E_{{\rm sc},x}}{dz^2}-\frac{E_{{\rm sc},x}}{\lambda^2}=0,
\end{equation}
where $\lambda$ is the London penetration depth. Taking into account the boundary conditions $E_{{\rm sc},x}|_{z=\pm d_{\rm f}/2}=E_0$, the solution of Eq.~(\ref{ec:London2}) in the SC regions is
\begin{equation}
\label{ec:E_sc}
E_{{\rm sc},x}(z)=E_0\,\exp\left(\frac{z\mp d_{\rm f}/2}{\lambda}\right),
\end{equation}
where the signs $-$ and $+$ correspond to the upper and lower SC layers, respectively. The absence of a dielectric spacer between SCs and the metallic FM allows current flow across the FM and closes the current loop. This prevents the formation of a nonuniform charge distribution in the SC and the related electric field, which could block SC current oscillations. It is worth mentioning that the in-plane resistance of the the FM layer is much higher than that of the SC layer. So, in our model, we are neglecting tangential currents in the FM layer. At the same time the normal resistance of the FM layer in our films is several order of magnitude lower than the tangential dynamic resistance of the SC layers. Thus, it is a good approximation to consider that the tangential currents are flowing along the superconducting layer, while in FM the currents are mainly normal.   

The alternating current in the SC layers can be obtained using the first London equation: 
\begin{equation}
\label{ec:London1}
E_{{\rm sc},x}=\frac{4\pi\lambda^2}{c^2}\frac{dj_x}{dt}.
\end{equation}
Here we note that the current density passing through the FM was neglected, since it is several orders of magnitude smaller than the $x$-component. This difference arises from the much larger cross-sectional area of the sample in the $xy$-plane compared to the $yz$-plane. The solution of Eq.~(\ref{ec:London1}) using Eqs.~(\ref{ec:E0}) and (\ref{ec:E_sc}) gives the following expression for the amplitude of the current density in the SC layers:
\begin{equation}
\label{ec:current}
j_{x}(z)=\mp \frac{d_{\rm f} c}{8 \pi \lambda^{2}} \left( h_{y}^{\rm sc} + 4 \pi m_{y} \right) 
\exp\left( \frac{z\mp d_{\rm f}/2}{\lambda} \right).
\end{equation}

Finally, the additional alternating magnetic field $h_y^{\rm sc}$ created by the SCs in the FM can be found using Ampere's law:
\begin{equation}
\label{ec:Maxwell2}
\oint\mathbf{h}\,d\mathbf{l}=\frac{4\pi}{c}\int\mathbf{j}\,d\mathbf{S}.
\end{equation} 

Let us consider a rectangular contour in the \(yz\)-plane, with side length \(L_y\) along the \(y\)-direction and height \(d_f/2 + d_s\) along the \(z\)-direction, chosen so that its sides run along the middle of the FM and the outer surface of the upper SC (Fig.~\ref{fig:SFS}) -- note that the static field $\mathbf{H}_0 = H_x \mathbf{e}_x$ can be omitted in (\ref{ec:Maxwell2}). Since in the top and bottom SC films current flows in opposite directions, the field cancels outside the SC/FM/SC structure and concentrates inside the films. Therefore, $h=h_y^{\rm sc}$ inside the FM, and $h=0$ for $z\geq d_{\rm f}/2+d_{\rm s}$. Applying Ampere's law, Eq.~(\ref{ec:Maxwell2}), to the top SC layer and taking into account the identical contribution to $h_y^{\rm sc}$ from the bottom SC layer, we can write: 
\begin{equation}
\label{ec:Max2_2}
h_y^{\rm sc}L_y=\frac{4\pi}{c}L_y\int_{d_{\rm f}/2}^{d_{\rm f}/2+d_{\rm s}} j_x(z)dz.
\end{equation} 
After substituting Eq.~(\ref{ec:current}) into Eq.~(\ref{ec:Max2_2}) and integrating the right-hand side, we obtain the expression for the amplitude $h_y^{\rm sc}$:
\begin{equation}
\label{ec:hy_sc}
h_y^{\rm sc}=-4\pi m_y\frac{d_{\rm f}\left[1 -\exp(-d_{\rm s}/\lambda) \right]}{2\lambda+d_{\rm f}\left[1 -\exp(-d_{\rm s}/\lambda) \right]}.
\end{equation} 
 
As a result, the field $h_y^{\rm sc}e^{i\omega t}$ should be included in the dynamical effective field in the Landau–Lifshitz equation. This modifies the Kittel formula (\ref{ec:omega}) and shifts the FMR frequency (field) upwards (downwards).

\section{\label{sec:level1} Summary}

We have studied the ferromagnetic resonance in a series of SC/FM/SC and SC/FM/I(M)/SC multilayers with varying thicknesses of both the superconducting and ferromagnetic layers. The temperature-dependent FMR measurements showed a strong reduction in the resonance field (i.e., an increase in the FMR frequency) at temperatures below the superconducting transition temperature of the SC layers. This effect was observed in the fully conductive Nb/Py/Nb and Nb/Py/Ag/Nb samples; adding a dielectric spacer between Nb and Py , as in Nb/Py/$\mathrm{Al_2O_3}$/Nb, made the effect vanish, however. Furthermore, for the SC/FM/SC and SC/FM/M/SC structures, the effect intensified with increasing the thickness of both the FM and SC layers and asymptotically saturation in the limits $d_f \to \infty$ or $d_s \to \infty$. The magnetostatic measurements showed essentially no change in hysteresis between the superconducting and normal states of Nb, indicating the dynamic origin of the observed effect.

We have developed an analytical model to explain the observed FMR frequency increase. In SC/FM/SC structures, the \emph{ac} magnetic field drives the magnetization precession in an FMR measurement, producing a time-varying magnetic flux in the structure. According to Faraday’s law, this results in the generation of an alternating electric current in the SC, which in turn produces an alternating magnetic field in the FM film, thereby shifting the FMR frequency. Crucially, this current arises only in the case of a conducting FM, in the absence of dielectric spacers, such that the screening-current loop is closed.

Our results clarify the mechanisms of the magnetization dynamics in SC/FM/SC multilayers, as well as address one of the key challenges in GHz magnonics, namely, increasing the operating frequency of magnonic devices.

\acknowledgments{\label{sec:level1} }
JWK acknowledge the financial support from the National Science Centre Poland: grants UMO-2020/39/O/ST5/02110, UMO-2021/43/I/ST3/00550. Support from the Swedish Research Council (VR~2018-03526), Olle Engkvist Foundation (2020-207-0460), Swedish Strategic Research Council (SSF UKR24-0002) are gratefully acknowledged.

\section*{Data Availability}
Data supporting this study are openly available from the repository at {https://doi.org/10.5281/zenodo.17235259}.


\begin{thebibliography}{27}%
\makeatletter
\providecommand \@ifxundefined [1]{%
 \@ifx{#1\undefined}
}%
\providecommand \@ifnum [1]{%
 \ifnum #1\expandafter \@firstoftwo
 \else \expandafter \@secondoftwo
 \fi
}%
\providecommand \@ifx [1]{%
 \ifx #1\expandafter \@firstoftwo
 \else \expandafter \@secondoftwo
 \fi
}%
\providecommand \natexlab [1]{#1}%
\providecommand \enquote  [1]{``#1''}%
\providecommand \bibnamefont  [1]{#1}%
\providecommand \bibfnamefont [1]{#1}%
\providecommand \citenamefont [1]{#1}%
\providecommand \href@noop [0]{\@secondoftwo}%
\providecommand \href [0]{\begingroup \@sanitize@url \@href}%
\providecommand \@href[1]{\@@startlink{#1}\@@href}%
\providecommand \@@href[1]{\endgroup#1\@@endlink}%
\providecommand \@sanitize@url [0]{\catcode `\\12\catcode `\$12\catcode `\&12\catcode `\#12\catcode `\^12\catcode `\_12\catcode `\%12\relax}%
\providecommand \@@startlink[1]{}%
\providecommand \@@endlink[0]{}%
\providecommand \url  [0]{\begingroup\@sanitize@url \@url }%
\providecommand \@url [1]{\endgroup\@href {#1}{\urlprefix }}%
\providecommand \urlprefix  [0]{URL }%
\providecommand \Eprint [0]{\href }%
\providecommand \doibase [0]{https://doi.org/}%
\providecommand \selectlanguage [0]{\@gobble}%
\providecommand \bibinfo  [0]{\@secondoftwo}%
\providecommand \bibfield  [0]{\@secondoftwo}%
\providecommand \translation [1]{[#1]}%
\providecommand \BibitemOpen [0]{}%
\providecommand \bibitemStop [0]{}%
\providecommand \bibitemNoStop [0]{.\EOS\space}%
\providecommand \EOS [0]{\spacefactor3000\relax}%
\providecommand \BibitemShut  [1]{\csname bibitem#1\endcsname}%
\let\auto@bib@innerbib\@empty
\bibitem [{\citenamefont {Yu}\ \emph {et~al.}(2025)\citenamefont {Yu}, \citenamefont {Zhou}, \citenamefont {Bauer},\ and\ \citenamefont {Bobkova}}]{yu2025}%
  \BibitemOpen
  \bibfield  {author} {\bibinfo {author} {\bibfnamefont {T.}~\bibnamefont {Yu}}, \bibinfo {author} {\bibfnamefont {X.-H.}\ \bibnamefont {Zhou}}, \bibinfo {author} {\bibfnamefont {G.~E.~W.}\ \bibnamefont {Bauer}},\ and\ \bibinfo {author} {\bibfnamefont {I.}~\bibnamefont {Bobkova}},\ }\href@noop {} {\bibinfo {title} {Electromagnetic proximity effect: Superconducting magnonics and beyond}} (\bibinfo {year} {2025}),\ \Eprint {https://arxiv.org/abs/2506.18502} {arXiv:2506.18502 [cond-mat.supr-con]} \BibitemShut {NoStop}%
\bibitem [{\citenamefont {Cai}\ \emph {et~al.}(2023)\citenamefont {Cai}, \citenamefont {Žutić},\ and\ \citenamefont {Han}}]{Cai_2022}%
  \BibitemOpen
  \bibfield  {author} {\bibinfo {author} {\bibfnamefont {R.}~\bibnamefont {Cai}}, \bibinfo {author} {\bibfnamefont {I.}~\bibnamefont {Žutić}},\ and\ \bibinfo {author} {\bibfnamefont {W.}~\bibnamefont {Han}},\ }\bibfield  {title} {\bibinfo {title} {Superconductor/ferromagnet heterostructures: A platform for superconducting spintronics and quantum computation},\ }\href {https://doi.org/https://doi.org/10.1002/qute.202200080} {\bibfield  {journal} {\bibinfo  {journal} {Adv. Quantum Technol.}\ }\textbf {\bibinfo {volume} {6}},\ \bibinfo {pages} {2200080} (\bibinfo {year} {2023})}\BibitemShut {NoStop}%
\bibitem [{\citenamefont {Bell}\ \emph {et~al.}(2008)\citenamefont {Bell}, \citenamefont {Milikisyants}, \citenamefont {Huber},\ and\ \citenamefont {Aarts}}]{Bell_2008}%
  \BibitemOpen
  \bibfield  {author} {\bibinfo {author} {\bibfnamefont {C.}~\bibnamefont {Bell}}, \bibinfo {author} {\bibfnamefont {S.}~\bibnamefont {Milikisyants}}, \bibinfo {author} {\bibfnamefont {M.}~\bibnamefont {Huber}},\ and\ \bibinfo {author} {\bibfnamefont {J.}~\bibnamefont {Aarts}},\ }\bibfield  {title} {\bibinfo {title} {Spin dynamics in a superconductor-ferromagnet proximity system},\ }\href {https://doi.org/10.1103/PhysRevLett.100.047002} {\bibfield  {journal} {\bibinfo  {journal} {Phys. Rev. Lett.}\ }\textbf {\bibinfo {volume} {100}},\ \bibinfo {pages} {047002} (\bibinfo {year} {2008})}\BibitemShut {NoStop}%
\bibitem [{\citenamefont {Jeon}\ \emph {et~al.}(2019)\citenamefont {Jeon}, \citenamefont {Ciccarelli}, \citenamefont {Kurebayashi}, \citenamefont {Cohen}, \citenamefont {Montiel}, \citenamefont {Eschrig}, \citenamefont {Wagner}, \citenamefont {Komori}, \citenamefont {Srivastava}, \citenamefont {Robinson},\ and\ \citenamefont {Blamire}}]{Jeon_2019}%
  \BibitemOpen
  \bibfield  {author} {\bibinfo {author} {\bibfnamefont {K.-R.}\ \bibnamefont {Jeon}}, \bibinfo {author} {\bibfnamefont {C.}~\bibnamefont {Ciccarelli}}, \bibinfo {author} {\bibfnamefont {H.}~\bibnamefont {Kurebayashi}}, \bibinfo {author} {\bibfnamefont {L.~F.}\ \bibnamefont {Cohen}}, \bibinfo {author} {\bibfnamefont {X.}~\bibnamefont {Montiel}}, \bibinfo {author} {\bibfnamefont {M.}~\bibnamefont {Eschrig}}, \bibinfo {author} {\bibfnamefont {T.}~\bibnamefont {Wagner}}, \bibinfo {author} {\bibfnamefont {S.}~\bibnamefont {Komori}}, \bibinfo {author} {\bibfnamefont {A.}~\bibnamefont {Srivastava}}, \bibinfo {author} {\bibfnamefont {J.~W.}\ \bibnamefont {Robinson}},\ and\ \bibinfo {author} {\bibfnamefont {M.~G.}\ \bibnamefont {Blamire}},\ }\bibfield  {title} {\bibinfo {title} {Effect of {Meissner} screening and trapped magnetic flux on magnetization dynamics in thick $\mathrm{Nb}/\mathrm{Ni}_{80}\mathrm{Fe}_{20}/\mathrm{Nb}$ trilayers},\ }\href {https://doi.org/10.1103/PhysRevApplied.11.014061} {\bibfield
  {journal} {\bibinfo  {journal} {Phys. Rev. Appl.}\ }\textbf {\bibinfo {volume} {11}},\ \bibinfo {pages} {014061} (\bibinfo {year} {2019})}\BibitemShut {NoStop}%
\bibitem [{\citenamefont {Golovchanskiy}\ \emph {et~al.}(2020)\citenamefont {Golovchanskiy}, \citenamefont {Abramov}, \citenamefont {Stolyarov}, \citenamefont {Chichkov}, \citenamefont {Silaev}, \citenamefont {Shchetinin}, \citenamefont {Golubov}, \citenamefont {Ryazanov}, \citenamefont {Ustinov},\ and\ \citenamefont {Kupriyanov}}]{Golovchanskiy_2020}%
  \BibitemOpen
  \bibfield  {author} {\bibinfo {author} {\bibfnamefont {I.}~\bibnamefont {Golovchanskiy}}, \bibinfo {author} {\bibfnamefont {N.}~\bibnamefont {Abramov}}, \bibinfo {author} {\bibfnamefont {V.}~\bibnamefont {Stolyarov}}, \bibinfo {author} {\bibfnamefont {V.}~\bibnamefont {Chichkov}}, \bibinfo {author} {\bibfnamefont {M.}~\bibnamefont {Silaev}}, \bibinfo {author} {\bibfnamefont {I.}~\bibnamefont {Shchetinin}}, \bibinfo {author} {\bibfnamefont {A.}~\bibnamefont {Golubov}}, \bibinfo {author} {\bibfnamefont {V.}~\bibnamefont {Ryazanov}}, \bibinfo {author} {\bibfnamefont {A.}~\bibnamefont {Ustinov}},\ and\ \bibinfo {author} {\bibfnamefont {M.}~\bibnamefont {Kupriyanov}},\ }\bibfield  {title} {\bibinfo {title} {Magnetization dynamics in proximity-coupled superconductor-ferromagnet-superconductor multilayers},\ }\href {https://doi.org/10.1103/PhysRevApplied.14.024086} {\bibfield  {journal} {\bibinfo  {journal} {Phys. Rev. Appl.}\ }\textbf {\bibinfo {volume} {14}},\ \bibinfo {pages} {024086} (\bibinfo {year}
  {2020})}\BibitemShut {NoStop}%
\bibitem [{\citenamefont {Golovchanskiy}\ \emph {et~al.}(2021)\citenamefont {Golovchanskiy}, \citenamefont {Abramov}, \citenamefont {Stolyarov}, \citenamefont {Weides}, \citenamefont {Ryazanov}, \citenamefont {Golubov}, \citenamefont {Ustinov},\ and\ \citenamefont {Kupriyanov}}]{Golovchanskiy_2021a}%
  \BibitemOpen
  \bibfield  {author} {\bibinfo {author} {\bibfnamefont {I.~A.}\ \bibnamefont {Golovchanskiy}}, \bibinfo {author} {\bibfnamefont {N.~N.}\ \bibnamefont {Abramov}}, \bibinfo {author} {\bibfnamefont {V.~S.}\ \bibnamefont {Stolyarov}}, \bibinfo {author} {\bibfnamefont {M.}~\bibnamefont {Weides}}, \bibinfo {author} {\bibfnamefont {V.~V.}\ \bibnamefont {Ryazanov}}, \bibinfo {author} {\bibfnamefont {A.~A.}\ \bibnamefont {Golubov}}, \bibinfo {author} {\bibfnamefont {A.~V.}\ \bibnamefont {Ustinov}},\ and\ \bibinfo {author} {\bibfnamefont {M.~Y.}\ \bibnamefont {Kupriyanov}},\ }\bibfield  {title} {\bibinfo {title} {Ultrastrong photon-to-magnon coupling in multilayered heterostructures involving superconducting coherence via ferromagnetic layers},\ }\href {https://doi.org/10.1126/sciadv.abe8638} {\bibfield  {journal} {\bibinfo  {journal} {Sci. Adv.}\ }\textbf {\bibinfo {volume} {7}},\ \bibinfo {pages} {eabe8638} (\bibinfo {year} {2021})}\BibitemShut {NoStop}%
\bibitem [{\citenamefont {Golovchanskiy}\ \emph {et~al.}(2023)\citenamefont {Golovchanskiy}, \citenamefont {Abramov}, \citenamefont {Emelyanova}, \citenamefont {Shchetinin}, \citenamefont {Ryazanov}, \citenamefont {Golubov},\ and\ \citenamefont {Stolyarov}}]{Golovchanskiy_2023}%
  \BibitemOpen
  \bibfield  {author} {\bibinfo {author} {\bibfnamefont {I.}~\bibnamefont {Golovchanskiy}}, \bibinfo {author} {\bibfnamefont {N.}~\bibnamefont {Abramov}}, \bibinfo {author} {\bibfnamefont {O.}~\bibnamefont {Emelyanova}}, \bibinfo {author} {\bibfnamefont {I.}~\bibnamefont {Shchetinin}}, \bibinfo {author} {\bibfnamefont {V.}~\bibnamefont {Ryazanov}}, \bibinfo {author} {\bibfnamefont {A.}~\bibnamefont {Golubov}},\ and\ \bibinfo {author} {\bibfnamefont {V.}~\bibnamefont {Stolyarov}},\ }\bibfield  {title} {\bibinfo {title} {Magnetization dynamics in proximity-coupled superconductor-ferromagnet-superconductor multilayers. ii. thickness dependence of the superconducting torque},\ }\href {https://doi.org/10.1103/PhysRevApplied.19.034025} {\bibfield  {journal} {\bibinfo  {journal} {Phys. Rev. Appl.}\ }\textbf {\bibinfo {volume} {19}},\ \bibinfo {pages} {034025} (\bibinfo {year} {2023})}\BibitemShut {NoStop}%
\bibitem [{\citenamefont {Li}\ \emph {et~al.}(2025)\citenamefont {Li}, \citenamefont {Zhang}, \citenamefont {Yang},\ and\ \citenamefont {Zhao}}]{Li2025EfficientFMR}%
  \BibitemOpen
  \bibfield  {author} {\bibinfo {author} {\bibfnamefont {J.}~\bibnamefont {Li}}, \bibinfo {author} {\bibfnamefont {J.}~\bibnamefont {Zhang}}, \bibinfo {author} {\bibfnamefont {G.}~\bibnamefont {Yang}},\ and\ \bibinfo {author} {\bibfnamefont {W.}~\bibnamefont {Zhao}},\ }\bibfield  {title} {\bibinfo {title} {Efficient shift of ferromagnetic resonance by superconductor gating},\ }\href {https://doi.org/10.1063/5.0231497} {\bibfield  {journal} {\bibinfo  {journal} {Appl. Phys. Rev.}\ }\textbf {\bibinfo {volume} {12}},\ \bibinfo {pages} {011419} (\bibinfo {year} {2025})}\BibitemShut {NoStop}%
\bibitem [{\citenamefont {Volkov}\ and\ \citenamefont {Efetov}(2009)}]{Volkov_2009}%
  \BibitemOpen
  \bibfield  {author} {\bibinfo {author} {\bibfnamefont {A.~F.}\ \bibnamefont {Volkov}}\ and\ \bibinfo {author} {\bibfnamefont {K.~B.}\ \bibnamefont {Efetov}},\ }\bibfield  {title} {\bibinfo {title} {Hybridization of spin and plasma waves in {Josephson} tunnel junctions containing a ferromagnetic layer},\ }\href {https://doi.org/10.1103/PhysRevLett.103.037003} {\bibfield  {journal} {\bibinfo  {journal} {Phys. Rev. Lett.}\ }\textbf {\bibinfo {volume} {103}},\ \bibinfo {pages} {037003} (\bibinfo {year} {2009})}\BibitemShut {NoStop}%
\bibitem [{\citenamefont {Mironov}\ and\ \citenamefont {Buzdin}(2021)}]{Mironov_2021}%
  \BibitemOpen
  \bibfield  {author} {\bibinfo {author} {\bibfnamefont {S.~V.}\ \bibnamefont {Mironov}}\ and\ \bibinfo {author} {\bibfnamefont {A.~I.}\ \bibnamefont {Buzdin}},\ }\bibfield  {title} {\bibinfo {title} {Giant demagnetization effects induced by superconducting films},\ }\href@noop {} {\bibfield  {journal} {\bibinfo  {journal} {J. Appl. Phys.}\ }\textbf {\bibinfo {volume} {119}},\ \bibinfo {pages} {102601} (\bibinfo {year} {2021})}\BibitemShut {NoStop}%
\bibitem [{\citenamefont {Silaev}(2022)}]{Silaev_2022}%
  \BibitemOpen
  \bibfield  {author} {\bibinfo {author} {\bibfnamefont {M.}~\bibnamefont {Silaev}},\ }\bibfield  {title} {\bibinfo {title} {Anderson-higgs mass of magnons in superconductor-ferromagnet-superconductor systems},\ }\href {https://doi.org/10.1103/PhysRevApplied.18.L061004} {\bibfield  {journal} {\bibinfo  {journal} {Phys. Rev. Appl.}\ }\textbf {\bibinfo {volume} {18}},\ \bibinfo {pages} {L061004} (\bibinfo {year} {2022})}\BibitemShut {NoStop}%
\bibitem [{\citenamefont {Ojaj\"arvi}\ \emph {et~al.}(2022)\citenamefont {Ojaj\"arvi}, \citenamefont {Bergeret}, \citenamefont {Silaev},\ and\ \citenamefont {Heikkil\"a}}]{Ojajarvi_2022}%
  \BibitemOpen
  \bibfield  {author} {\bibinfo {author} {\bibfnamefont {R.}~\bibnamefont {Ojaj\"arvi}}, \bibinfo {author} {\bibfnamefont {F.~S.}\ \bibnamefont {Bergeret}}, \bibinfo {author} {\bibfnamefont {M.~A.}\ \bibnamefont {Silaev}},\ and\ \bibinfo {author} {\bibfnamefont {T.~T.}\ \bibnamefont {Heikkil\"a}},\ }\bibfield  {title} {\bibinfo {title} {Dynamics of two ferromagnetic insulators coupled by superconducting spin current},\ }\href {https://doi.org/10.1103/PhysRevLett.128.167701} {\bibfield  {journal} {\bibinfo  {journal} {Phys. Rev. Lett.}\ }\textbf {\bibinfo {volume} {128}},\ \bibinfo {pages} {167701} (\bibinfo {year} {2022})}\BibitemShut {NoStop}%
\bibitem [{\citenamefont {Kuznetsov}\ and\ \citenamefont {Fraerman}(2022)}]{Kuznetsov_2022}%
  \BibitemOpen
  \bibfield  {author} {\bibinfo {author} {\bibfnamefont {M.~A.}\ \bibnamefont {Kuznetsov}}\ and\ \bibinfo {author} {\bibfnamefont {A.~A.}\ \bibnamefont {Fraerman}},\ }\bibfield  {title} {\bibinfo {title} {Temperature-sensitive spin-wave nonreciprocity induced by interlayer dipolar coupling in ferromagnet/paramagnet and ferromagnet/superconductor hybrid systems},\ }\href {https://doi.org/10.1103/PhysRevB.105.214401} {\bibfield  {journal} {\bibinfo  {journal} {Phys. Rev. B}\ }\textbf {\bibinfo {volume} {105}},\ \bibinfo {pages} {214401} (\bibinfo {year} {2022})}\BibitemShut {NoStop}%
\bibitem [{\citenamefont {Silaev}(2023)}]{Silaev_2023}%
  \BibitemOpen
  \bibfield  {author} {\bibinfo {author} {\bibfnamefont {M.}~\bibnamefont {Silaev}},\ }\bibfield  {title} {\bibinfo {title} {Ultrastrong magnon-photon coupling, squeezed vacuum, and entanglement in superconductor/ferromagnet nanostructures},\ }\href {https://doi.org/10.1103/PhysRevB.107.L180503} {\bibfield  {journal} {\bibinfo  {journal} {Phys. Rev. B}\ }\textbf {\bibinfo {volume} {107}},\ \bibinfo {pages} {L180503} (\bibinfo {year} {2023})}\BibitemShut {NoStop}%
\bibitem [{\citenamefont {Zhou}\ and\ \citenamefont {Yu}(2023)}]{Zhou_2023}%
  \BibitemOpen
  \bibfield  {author} {\bibinfo {author} {\bibfnamefont {X.-H.}\ \bibnamefont {Zhou}}\ and\ \bibinfo {author} {\bibfnamefont {T.}~\bibnamefont {Yu}},\ }\bibfield  {title} {\bibinfo {title} {Gating ferromagnetic resonance of magnetic insulators by superconductors via modulating electric field radiation},\ }\href {https://doi.org/10.1103/PhysRevB.108.144405} {\bibfield  {journal} {\bibinfo  {journal} {Phys. Rev. B}\ }\textbf {\bibinfo {volume} {108}},\ \bibinfo {pages} {144405} (\bibinfo {year} {2023})}\BibitemShut {NoStop}%
\bibitem [{\citenamefont {Zhou}\ \emph {et~al.}(2024)\citenamefont {Zhou}, \citenamefont {Ye}, \citenamefont {Bai},\ and\ \citenamefont {Yu}}]{Zhou_2024}%
  \BibitemOpen
  \bibfield  {author} {\bibinfo {author} {\bibfnamefont {X.-H.}\ \bibnamefont {Zhou}}, \bibinfo {author} {\bibfnamefont {X.}~\bibnamefont {Ye}}, \bibinfo {author} {\bibfnamefont {L.}~\bibnamefont {Bai}},\ and\ \bibinfo {author} {\bibfnamefont {T.}~\bibnamefont {Yu}},\ }\bibfield  {title} {\bibinfo {title} {Giant enhancement of magnon transport by superconductor {Meissner} screening},\ }\href {https://doi.org/10.1103/PhysRevB.110.L020404} {\bibfield  {journal} {\bibinfo  {journal} {Phys. Rev. B}\ }\textbf {\bibinfo {volume} {110}},\ \bibinfo {pages} {L020404} (\bibinfo {year} {2024})}\BibitemShut {NoStop}%
\bibitem [{\citenamefont {Derendorf}\ \emph {et~al.}(2024)\citenamefont {Derendorf}, \citenamefont {Volkov},\ and\ \citenamefont {Eremin}}]{Derendorf_2024}%
  \BibitemOpen
  \bibfield  {author} {\bibinfo {author} {\bibfnamefont {P.}~\bibnamefont {Derendorf}}, \bibinfo {author} {\bibfnamefont {A.~F.}\ \bibnamefont {Volkov}},\ and\ \bibinfo {author} {\bibfnamefont {I.~M.}\ \bibnamefont {Eremin}},\ }\bibfield  {title} {\bibinfo {title} {Superluminal propagation of composite collective modes in superconductor-ferromagnet heterostructures},\ }\href {https://doi.org/10.1103/PhysRevB.110.144502} {\bibfield  {journal} {\bibinfo  {journal} {Phys. Rev. B}\ }\textbf {\bibinfo {volume} {110}},\ \bibinfo {pages} {144502} (\bibinfo {year} {2024})}\BibitemShut {NoStop}%
\bibitem [{\citenamefont {Gordeeva}\ \emph {et~al.}(2025)\citenamefont {Gordeeva}, \citenamefont {Bobkov}, \citenamefont {Bobkov}, \citenamefont {Bobkova},\ and\ \citenamefont {Yu}}]{Gordeeva_2025}%
  \BibitemOpen
  \bibfield  {author} {\bibinfo {author} {\bibfnamefont {V.~M.}\ \bibnamefont {Gordeeva}}, \bibinfo {author} {\bibfnamefont {G.~A.}\ \bibnamefont {Bobkov}}, \bibinfo {author} {\bibfnamefont {A.~M.}\ \bibnamefont {Bobkov}}, \bibinfo {author} {\bibfnamefont {I.~V.}\ \bibnamefont {Bobkova}},\ and\ \bibinfo {author} {\bibfnamefont {T.}~\bibnamefont {Yu}},\ }\bibfield  {title} {\bibinfo {title} {Ultrastrong coupling of two ferromagnets via {Meissner} currents},\ }\href {https://doi.org/10.1103/j447-3jl4} {\bibfield  {journal} {\bibinfo  {journal} {Phys. Rev. B}\ }\textbf {\bibinfo {volume} {111}},\ \bibinfo {pages} {214520} (\bibinfo {year} {2025})}\BibitemShut {NoStop}%
\bibitem [{\citenamefont {Seshadri}(1970)}]{Seshadri_1970}%
  \BibitemOpen
  \bibfield  {author} {\bibinfo {author} {\bibfnamefont {S.}~\bibnamefont {Seshadri}},\ }\bibfield  {title} {\bibinfo {title} {Surface magnetostatic modes of a ferrite slab},\ }\href {https://doi.org/10.1109/PROC.1970.7680} {\bibfield  {journal} {\bibinfo  {journal} {Proc. IEEE}\ }\textbf {\bibinfo {volume} {58}},\ \bibinfo {pages} {506} (\bibinfo {year} {1970})}\BibitemShut {NoStop}%
\bibitem [{\citenamefont {Borst}\ \emph {et~al.}(2023)\citenamefont {Borst}, \citenamefont {Vree}, \citenamefont {Lowther}, \citenamefont {Teepe}, \citenamefont {Kurdi}, \citenamefont {Bertelli}, \citenamefont {Simon}, \citenamefont {Blanter},\ and\ \citenamefont {van~der Sar}}]{Borst_2023}%
  \BibitemOpen
  \bibfield  {author} {\bibinfo {author} {\bibfnamefont {M.}~\bibnamefont {Borst}}, \bibinfo {author} {\bibfnamefont {P.~H.}\ \bibnamefont {Vree}}, \bibinfo {author} {\bibfnamefont {A.}~\bibnamefont {Lowther}}, \bibinfo {author} {\bibfnamefont {A.}~\bibnamefont {Teepe}}, \bibinfo {author} {\bibfnamefont {S.}~\bibnamefont {Kurdi}}, \bibinfo {author} {\bibfnamefont {I.}~\bibnamefont {Bertelli}}, \bibinfo {author} {\bibfnamefont {B.~G.}\ \bibnamefont {Simon}}, \bibinfo {author} {\bibfnamefont {Y.~M.}\ \bibnamefont {Blanter}},\ and\ \bibinfo {author} {\bibfnamefont {T.}~\bibnamefont {van~der Sar}},\ }\bibfield  {title} {\bibinfo {title} {Observation and control of hybrid spin-wave–{Meissner}-current transport modes},\ }\href {https://doi.org/10.1126/science.adj7576} {\bibfield  {journal} {\bibinfo  {journal} {Science}\ }\textbf {\bibinfo {volume} {382}},\ \bibinfo {pages} {430} (\bibinfo {year} {2023})}\BibitemShut {NoStop}%
\bibitem [{\citenamefont {Swihart}(1961)}]{Swihart_1961}%
  \BibitemOpen
  \bibfield  {author} {\bibinfo {author} {\bibfnamefont {J.~C.}\ \bibnamefont {Swihart}},\ }\bibfield  {title} {\bibinfo {title} {Field solution for a thin‐film superconducting strip transmission line},\ }\href {https://doi.org/10.1063/1.1736025} {\bibfield  {journal} {\bibinfo  {journal} {J. Appl. Phys.}\ }\textbf {\bibinfo {volume} {32}},\ \bibinfo {pages} {461} (\bibinfo {year} {1961})}\BibitemShut {NoStop}%
\bibitem [{\citenamefont {Tserkovnyak}\ \emph {et~al.}(2005)\citenamefont {Tserkovnyak}, \citenamefont {Brataas}, \citenamefont {Bauer},\ and\ \citenamefont {Halperin}}]{Tserkovnyak_2005}%
  \BibitemOpen
  \bibfield  {author} {\bibinfo {author} {\bibfnamefont {Y.}~\bibnamefont {Tserkovnyak}}, \bibinfo {author} {\bibfnamefont {A.}~\bibnamefont {Brataas}}, \bibinfo {author} {\bibfnamefont {G.~E.~W.}\ \bibnamefont {Bauer}},\ and\ \bibinfo {author} {\bibfnamefont {B.~I.}\ \bibnamefont {Halperin}},\ }\bibfield  {title} {\bibinfo {title} {Nonlocal magnetization dynamics in ferromagnetic heterostructures},\ }\href {https://doi.org/10.1103/RevModPhys.77.1375} {\bibfield  {journal} {\bibinfo  {journal} {Rev. Mod. Phys.}\ }\textbf {\bibinfo {volume} {77}},\ \bibinfo {pages} {1375} (\bibinfo {year} {2005})}\BibitemShut {NoStop}%
\bibitem [{\citenamefont {Deutscher}\ and\ \citenamefont {Feinberg}(2000)}]{Deutscher_2000}%
  \BibitemOpen
  \bibfield  {author} {\bibinfo {author} {\bibfnamefont {G.}~\bibnamefont {Deutscher}}\ and\ \bibinfo {author} {\bibfnamefont {D.}~\bibnamefont {Feinberg}},\ }\bibfield  {title} {\bibinfo {title} {Coupling superconducting-ferromagnetic point contacts by andreev reflections},\ }\href {https://doi.org/10.1063/1.125796} {\bibfield  {journal} {\bibinfo  {journal} {Appl. Phys. Lett.}\ }\textbf {\bibinfo {volume} {76}},\ \bibinfo {pages} {487} (\bibinfo {year} {2000})}\BibitemShut {NoStop}%
\bibitem [{\citenamefont {Keizer}\ \emph {et~al.}()\citenamefont {Keizer}, \citenamefont {Goennenwein}, \citenamefont {Klapwijk}, \citenamefont {Miao}, \citenamefont {Xiao},\ and\ \citenamefont {Gupta}}]{Keizer_2006}%
  \BibitemOpen
  \bibfield  {author} {\bibinfo {author} {\bibfnamefont {R.~S.}\ \bibnamefont {Keizer}}, \bibinfo {author} {\bibfnamefont {S.~T.~B.}\ \bibnamefont {Goennenwein}}, \bibinfo {author} {\bibfnamefont {T.~M.}\ \bibnamefont {Klapwijk}}, \bibinfo {author} {\bibfnamefont {G.}~\bibnamefont {Miao}}, \bibinfo {author} {\bibfnamefont {G.}~\bibnamefont {Xiao}},\ and\ \bibinfo {author} {\bibfnamefont {A.}~\bibnamefont {Gupta}},\ }\bibfield  {title} {\bibinfo {title} {A spin triplet supercurrent through the half-metallic ferromagnet {CrO$_2$}},\ }\href {https://doi.org/10.1038/nature04499} {\bibfield  {journal} {\bibinfo  {journal} {Nature}\ }\textbf {\bibinfo {volume} {439}},\ \bibinfo {pages} {825}}\BibitemShut {NoStop}%
\bibitem [{\citenamefont {Li}\ \emph {et~al.}(2018)\citenamefont {Li}, \citenamefont {Y.-L.Zhao}, \citenamefont {Zhang},\ and\ \citenamefont {Sun}}]{Li}%
  \BibitemOpen
  \bibfield  {author} {\bibinfo {author} {\bibfnamefont {L.-L.}\ \bibnamefont {Li}}, \bibinfo {author} {\bibnamefont {Y.-L.Zhao}}, \bibinfo {author} {\bibfnamefont {X.-X.}\ \bibnamefont {Zhang}},\ and\ \bibinfo {author} {\bibfnamefont {Y.}~\bibnamefont {Sun}},\ }\bibfield  {title} {\bibinfo {title} {Possible evidence for spin-transfer torque induced by spin-triplet supercurrents},\ }\href@noop {} {\bibfield  {journal} {\bibinfo  {journal} {Chin. Phys. Lett.}\ }\textbf {\bibinfo {volume} {35}},\ \bibinfo {pages} {077401} (\bibinfo {year} {2018})},\ \translation{Chin. Phys. Lett. \textbf{35}, 077401 (2018)}\BibitemShut {NoStop}%
\bibitem [{\citenamefont {Gubin}\ \emph {et~al.}(2005)\citenamefont {Gubin}, \citenamefont {Il'in}, \citenamefont {Vitusevich}, \citenamefont {Siegel},\ and\ \citenamefont {Klein}}]{Gubin}%
  \BibitemOpen
  \bibfield  {author} {\bibinfo {author} {\bibfnamefont {A.~I.}\ \bibnamefont {Gubin}}, \bibinfo {author} {\bibfnamefont {K.~S.}\ \bibnamefont {Il'in}}, \bibinfo {author} {\bibfnamefont {S.~A.}\ \bibnamefont {Vitusevich}}, \bibinfo {author} {\bibfnamefont {M.}~\bibnamefont {Siegel}},\ and\ \bibinfo {author} {\bibfnamefont {N.}~\bibnamefont {Klein}},\ }\bibfield  {title} {\bibinfo {title} {Dependence of magnetic penetration depth on the thickness of superconducting {Nb} thin films},\ }\href {https://doi.org/10.1103/PhysRevB.72.064503} {\bibfield  {journal} {\bibinfo  {journal} {Phys. Rev. B}\ }\textbf {\bibinfo {volume} {72}},\ \bibinfo {pages} {064503} (\bibinfo {year} {2005})}\BibitemShut {NoStop}%
\bibitem [{\citenamefont {Tinkham}(1996)}]{Tinkham1996}%
  \BibitemOpen
  \bibfield  {author} {\bibinfo {author} {\bibfnamefont {M.}~\bibnamefont {Tinkham}},\ }\href@noop {} {\emph {\bibinfo {title} {Introduction to Superconductivity}}},\ \bibinfo {edition} {2nd}\ ed.\ (\bibinfo  {publisher} {McGraw-Hill},\ \bibinfo {address} {New York},\ \bibinfo {year} {1996})\BibitemShut {NoStop}%
\end{thebibliography}

\providecommand{\noopsort}[1]{}\providecommand{\singleletter}[1]{#1}%

\end{document}